\documentclass[letterpaper, 10 pt, conference]{ieeeconf}  

\IEEEoverridecommandlockouts                              
\usepackage{graphicx}      
\usepackage{epstopdf}
\usepackage{amsmath}
\usepackage{algcompatible}
\usepackage{amssymb}
\usepackage{enumerate}
\usepackage[usenames, dvipsnames]{color}
\usepackage{algorithm}
\usepackage{algpseudocode}
\usepackage{mathtools}
\usepackage{breqn}
\usepackage{bbm}
 \usepackage{cite}
\graphicspath{{./figures/}}
\newcommand{\REAL}{\ensuremath{\mathbb{R}}}

\newcommand{\beq}{\begin{equation}}
\newcommand{\eeq}{\end{equation}}
\newcommand{\bea}{\begin{eqnarray}}
\newcommand{\eea}{\end{eqnarray}}
\newcommand{\bean}{\begin{eqnarray*}}
\newcommand{\eean}{\end{eqnarray*}}
\newcommand{\bcen}{\begin{center}}
\newcommand{\ecen}{\end{center}}
\newcommand{\bitm}{\begin{itemize}}
\newcommand{\eitm}{\end{itemize}}

\def\sgcnv{\hbox{$\, \bigcirc \,$\kern-0.9em\hbox{\mgop}$\,$}} 
\def\supgeno{\hbox{$\, \bigcirc \,$\kern-1.0em\hbox{$\wedge$}$\,$}} 
\def\infgeno{\hbox{$\, \bigcirc \,$\kern-1.0em\hbox{$\vee$}$\,$}} 
\newcommand{\mgop}{\ensuremath{\star}}  

\usepackage{pdfpages}
\newtheorem{definition}{\bf Definition}
\newtheorem{assumption}{\bf Assumption}
\newtheorem{remark}{\bf Remark}
\newtheorem{theorem}{\bf Theorem}
\newtheorem{lemma}{\bf Lemma}
\newtheorem{proposition}{\bf Proposition}

\newcommand{\abs}[1]{\ensuremath{\left\vert #1\right\vert}}

\newcommand{\paren}[1]{\ensuremath{\left( #1\right)}}
\newcommand{\clint}[1]{\ensuremath{\left[ #1\right]}}

\newcommand{\set}[1]{\ensuremath{\left\{ #1\right\}}}
\newcommand{\vct}[1]{\ensuremath{\boldsymbol{#1}}}  

\newcommand{\matr}[1]{\ensuremath{\clint{\begin{array} #1 \end{array}}}}
\newcommand{\expe}[1]{\ensuremath{\mathbb{E}\set{#1}}}

\newcommand{\var}[1]{\ensuremath{\mathrm{Cov}\set{#1}}}

\newcommand{\info}[1]{\ensuremath{\mathcal{I}_{#1}}}

\newcommand{\event}{\ensuremath{\mathcal{B}_{k_0}}}

\newcommand{\y}[1]{\ensuremath{z_{#1}}}

\usepackage{array}
\newcolumntype{"}{@{\hskip\tabcolsep\vrule width 2pt\hskip\tabcolsep}}

\begin{document}

\title{An Information Matrix Approach for State Secrecy}

\author{Anastasios~Tsiamis, Konstantinos~Gatsis, George~J.~Pappas
\thanks{This work was supported in part by ONR N00014-17-1-2012, and by NSF CNS-1505799 grant and the Intel-NSF Partnership for Cyber-Physical Systems Security and Privacy. }
\thanks{The authors  are   with   the   Department   of   Electrical   and   Systems  Engineering,  University  of  Pennsylvania,  Philadelphia,  PA  19104.
		 Emails: \{atsiamis,kgatsis,pappasg\}@seas.upenn.edu
}
}
\maketitle
\begin{abstract}
This paper studies the problem of remote state estimation in the presence of a passive eavesdropper. A sensor measures a linear plant's state and transmits it to an authorized user over a packet-dropping channel, which is susceptible to eavesdropping. 
Our goal is to design a coding scheme such that the eavesdropper cannot infer the plant's current state, while the user successfully decodes the sent messages.
We employ a novel class of codes, termed State-Secrecy Codes, which are fast and efficient for dynamical systems. They apply linear time-varying transformations to the current and past states received by the user. In this way, they force the eavesdropper's information matrix  to decrease with asymptotically the same rate as in the open-loop prediction case, i.e. when the eavesdropper misses all messages. As a result, the eavesdropper's minimum mean square error (mmse) for the unstable states grows unbounded, while the respective error for the stable states converges to the open-loop prediction one. These secrecy guarantees are achieved under minimal conditions, which require that, at least once, the user receives the corresponding packet while the eavesdropper fails to intercept it.  
Meanwhile, the user's estimation performance remains optimal. 
The theoretical results are illustrated in simulations. 
\end{abstract}

\IEEEpeerreviewmaketitle

\section{Introduction}\label{Section_Introduction}
In this paper, we study passive eavesdropping attacks in a remote estimation setting. This scenario represents Internet of Things applications, where sensors collect confidential information
about the state of a dynamical system and send it to an
authorized user, e.g. a controller, a cloud server, etc., through
a wireless channel. Due to the broadcast nature of the wireless medium, this confidential information might get leaked to eavesdroppers~\cite{Survey_Wireless_Security}. Our goal is to design codes such that the authorized user can estimate the state of the plant, while any eavesdroppers eventually lose track of the state. 
We only deal with eavesdropping attacks here, but other types of attacks have also been studied~\cite{sandberg2015cyberphysical}. Those include denial-of-service attacks~\cite{gupta2010optimal} and data-integrity attacks~\cite{mo2014resilient,fawzi2014secure,pajic2017attack,pasqualetti2015control,lesi2017security}. 

One of the main challenges when designing codes for secret communications is the tradeoff between code complexity and security. 
Encryption methods~\cite{Cryptography} offer confidentiality guarantees without requiring any mathematical model of the physical components, i.e. the source or the channel. They might introduce computational and communication overheads~\cite{lee2010price} though, and their effectiveness is based on the assumption that the adversaries are computationally bounded. 
A question that naturally arises is whether we can incorporate model knowledge in order to develop additional defenses.

Information theoretic approaches develop codes in the physical layer of wireless communications by exploiting the channel model \cite{Proc_IEEE_Special_issue, Wyner_wiretap,LiangPoor2008Secure,li2011communication,wiese2016secure}. 
 The provided secrecy guarantees are provable, strong and are independent of the eavesdropper's computational capability. Constructing such codes is challenging and requires knowledge of the eavesdropper's channel model.
In the case of packet erasure channels, more practical codes can be designed~\cite{safaka2016creating}.

In the case of dynamical systems, the dynamics provide an additional structure that could be exploited for secrecy.
In this work, we generalize State-Secrecy Codes, a new class of codes for linear systems, which indeed exploit the dynamics and the channel randomness.  The current state is encoded by subtracting from it a weighted version of the user's previously received state.  This operation has low complexity and requires acknowledgment signals from the user back to the sensor.  
Under minimal conditions on the communication channel, which is modeled as a packet dropping one, the eavesdropper's information matrix (inverse mmse covariance matrix) converges to the open-loop prediction one, i.e. the information matrix when the eavesdropper misses all messages. This is because our code introduces artificial dynamics to the eavesdropper's information matrix recursion, forcing it to decrease with asymptotically the same rate as in the open-loop case (see Remarks~\ref{REM_artificial_dynamics},~\ref{REM_Weighting_Matrix_Form} in Section~\ref{Section_Solution}).
As a result, the eavesdropper's mmse for the unstable states diverges to infinity, while the mmse for the stable states converges to the open-loop one (Theorem~\ref{THM_perfect_secrecy} in Section~\ref{Section_Solution}). 
 The channel conditions only require that at least once the user receives the corresponding packet while the eavesdropper misses it. Meanwhile, the user can always decode the packets and has optimal mmse. 

Related work can be found in~\cite{leong2017ifac,leong2017MarkovianSecrecy,tsiamis2017state}, where a non-coding approach is adopted. A simple mechanism which withholds measurements is employed, but with high probability the eavesdropper might have very small estimation error infinitely often.
 Preliminary versions of our scheme appeared in~\cite{tsiamis2017codes,tsiamis2017codes_stable}, but the results are limited to protecting either unstable states~\cite{tsiamis2017codes} or purely stable systems~\cite{tsiamis2017codes_stable}. In this paper, we develop a new unified framework for general linear systems, based on a novel analysis from the point of view of information matrices. The coding schemes in~\cite{tsiamis2017codes,tsiamis2017codes_stable} can be obtained as special cases of the present scheme. The converse is not true as illustrated in Section~\ref{Section_Simulations}.

Designed specifically for dynamical systems, State-Secrecy Codes offer a good tradeoff between code complexity and secrecy of the current state:
\begin{itemize}
	\item They are simpler than information theoretic codes and encryption, they do not require knowledge of the eavesdropper's channel and they avoid communication overheads.
	\item The confidentiality guarantees for the plant's current state are comparable to the information theoretic ones, overcoming the limitations of~\cite{leong2017ifac,leong2017MarkovianSecrecy,tsiamis2017state}; almost surely the eavesdropper's information matrix converges to the open-loop prediction one. These guarantees do not depend on the computational capabilities of the eavesdropper.
\end{itemize}

\section{Problem formulation}\label{Section_Formulation}
\begin{figure}[t] \centering{
		\includegraphics[scale=0.75]{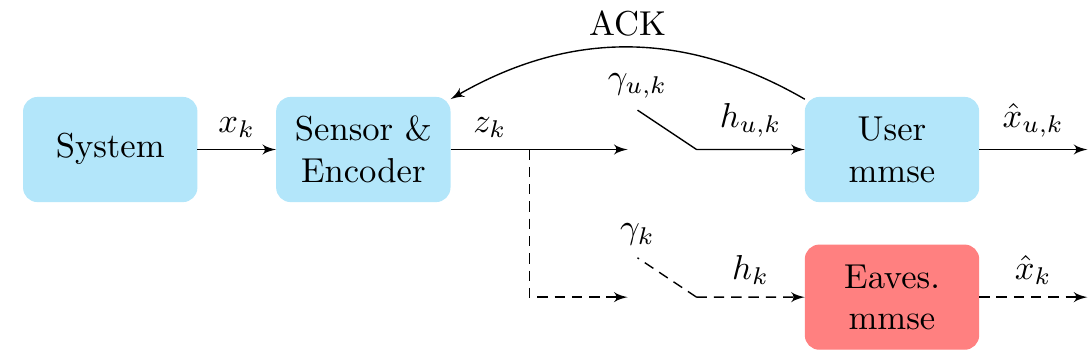}  
		\caption {A sensor collects the state $x_k$ of the dynamical system \eqref{EQN_system} and transmits $z_{k}$, an encoded version of the state, over a packet dropping channel. The packets might be dropped by the authorized user, as captured by $\gamma_{u,k}$, and might be intercepted by the eavesdropper, as captured by $\gamma_{k}$. Both the user and the eavesdropper compute the minimum mean square error (mmse) estimates  $\hat{x}_{u,k}$ and $\hat{x}_{k}$ respectively. The user transmits acknowledgment signals back to the sensor.
		}
		\label{Figure_ProblemSetup} }
\end{figure}

The considered remote estimation architecture consists of a sensor observing a dynamical system, a packet dropping channel, a legitimate user, and an eavesdropper--see Figure~\ref{Figure_ProblemSetup}. 
\subsection{Dynamical system model}
 The dynamical system is modeled as discrete-time linear:
\begin{align} \label{EQN_system}
x_{k+1}&=Ax_{k}+w_{k+1},
\end{align}
where $x_{k}\in \REAL^{n}$ is the state, $A\in\REAL^{n\times n}$ is the system matrix, and $w_{k}\in \REAL^{n}$ is the process noise, modeled as i.i.d. Gaussian with zero mean and covariance $Q$. The initial state $x_{0}$ is also Gaussian with zero mean, covariance $\Sigma_0$ and is independent of the process noise. 
All system and noise parameters $A, Q,\Sigma_0$  are assumed to be public knowledge, available to all involved entities, i.e., the sensor, the user, and the eavesdropper. 
 The following assumptions hold throughout this paper.
\begin{assumption}\label{ASSUM_system}
Matrix $A$ in~\eqref{EQN_system} has no eigenvalues on the unit circle and is invertible. Matrices $Q,\,\Sigma_0$ are 
positive definite: $Q,\,\Sigma_0\succ 0$, where $\succ$ ($\succeq$) denotes comparison in the positive definite (semidefinite) cone.  \hfill $\diamond$
\end{assumption}
The case of eigenvalues on the unit circle or zero eigenvalues is discussed in Section~\ref{Section_Discussion}. 
Without loss of generality, we can assume the system is in real Jordan form (see~\cite{horn2012matrix} ch. 3.4).  \begin{assumption}\label{ASSUM_Jordan}
	Matrix $A$ is in real Jordan form:
	\begin{equation}\label{EQN_subsystems}
	A=\matr{{cc}A_u&0\\0&A_s},
	\end{equation}
	where $A_u\in\REAL^{n_{u}\times n_{u}}$ is the Jordan form of the unstable part and $A_s\in\REAL^{n_s\times n_s}$ is the Jordan form of the stable part.\hfill $\diamond$
\end{assumption}
We denote by $\abs{\lambda_{i}}\in \mathbb{R}$ the magnitude of the eigenvalue of $A$ that corresponds to the block of $A_{ii}$. Matrix $Q$ in block form is written as \[Q=\matr{{cc}Q_{u}&Q_{12}\\Q_{12}'&Q_{s}}.\]

\subsection{Channel model}
Communication follows the packet-based paradigm commonly used in networked control systems \cite{Sinopoli2004Kalman,hespanha2007survey,GatsisEtal14}. The sensor transmits encoded state information $z_k\in \REAL^n$ over a packet dropping channel with two outputs/receivers as shown in Figure~\ref{Figure_ProblemSetup}.
The first output, denoted by $h_{u,k}$, is the authorized one to the user, while the second, denoted by $h_{k}$, is the unauthorized one to the eavesdropper.

Communication with the user is unreliable, i.e. packets might be dropped. Respectively, communication is not secure against the eavesdropper, i.e., the latter may intercept transmitted packets. 
We denote by $\gamma_{u,k}\in\set{0,1}$ the outcome of the user packet reception at time $k$, and by $\gamma_{k}\in\set{0,1}$ the outcome of the eavesdropper's packet interception. If $\gamma_{u,k}=1$ (or $\gamma_{k}=1$), then the reception (interception) is successful. Otherwise, the respective packet is dropped.  The outputs of the channel are modeled as:
\begin{equation}\label{EQN_channel_model} 
h_{u,k}=\left\{
\begin{aligned}\y{k}, \quad &\text{ if } \gamma_{u,k}=1\\
\varepsilon,  \quad &\text{ if } \gamma_{u,k}=0
\end{aligned}\right.,
h_{k}=\left\{
\begin{aligned}\y{k}, \quad &\text{ if } \gamma_{k}=1\\
\varepsilon,  \quad &\text{ if } \gamma_{k}=0
\end{aligned}\right.
\end{equation}
where symbol $\varepsilon$, is used to represent the ``no information" outcome. The channel outcomes $\{\gamma_{u,k},\gamma_{k}, k\ge 0\}$ are assumed to be independent of the initial state $x_0$, and the process noise $w_k$, for $k\ge 0$. No specific joint distribution of the channel outcomes is assumed.

In addition to the main channel, the user can reliably send acknowledgment signals back to the sensor via the reverse channel.
The case of unreliable acknowledgments is discussed in Section~\ref{Section_Discussion}. 
Thus, at any time step the sensor knows what is the latest received message $\y{k}$ at the user. 
Meanwhile, the eavesdropper is able to intercept all acknowledgment signals and knows the history of user's packet successes. In that respect, we model a powerful eavesdropper. 
Neither the sensor nor the user have any knowledge about the eavesdropper's intercept successes $\gamma_{k}$. 

\subsection{MMSE Estimation}
Both the user and the eavesdropper know the encoding scheme
and use a minimum mean square error (mmse) estimate to decode the received/intercepted messages. 
This estimate depends on their information up to time $k$. We define the user's batch vector of channel outputs by $\vct{h}_{u,0:k}=\paren{h_{u,0}, \ldots, h_{u,k}}$ and the batch vector of channel outcomes by $\vct{\gamma}_{u,0:k}=\paren{\gamma_{u,0},\dots,\gamma_{u,k}}$. The eavesdropper's batch vectors $\vct{h}_{0:k},\,\vct{\gamma}_{0:k}$ are defined similarly.
Then, the user's information at time $k$ is denoted by $\mathcal{I}^u_k=\{\vct{h}_{u,0:k} \}$, with $\mathcal{I}^u_{-1}=\emptyset$. Respectively, we denote the eavesdropper's information by
\begin{equation}\label{EQN_eavesdropper_information}
\info k=\set{\vct{h}_{0:k},\vct{\gamma}_{u,0:k}},\,\info{-1}=\emptyset
\end{equation} 
Notice that the eavesdropper has the additional information of the user's reception success history.
The eavesdropper's mmse estimate, $\hat{x}_{k}$, and the respective mmse covariance matrix $P_{k}$ are given by:
\begin{align}\label{EQN_estimate_variance_eavesdropper}
\hat{x}_{k}=\expe{ x_{k}\vert \info k},\quad P_{k}=\var{x_{k}\vert\info k},
\end{align}
where $\var{x_{k}|\info k}=\expe{\paren{x_{k}-\hat{x}_{k}}\paren{x_{k}-\hat{x}_{k}}'|\info k}$.
The user's mmse estimate, $\hat{x}_{u,k}$ and the respective mmse covariance matrix $P_{u,k}$ are defined similarly.

\subsection{Problem}
The goal of this work is to design a coding scheme at the sensor, so that we achieve \emph{perfect secrecy} for the current state (introduced below). 
We require the user's estimation scheme to be optimal,  i.e. to have zero estimation error at the successful reception times. At the same time, we require the eavesdropper's mmse error to behave asymptotically as in open-loop prediction case, i.e., when the eavesdropper misses all signals $z_k$. The motivation is that the open-loop prediction mmse is maximum in expectation (see~\cite{tsiamisJournal}); in that aspect, it represents the worst case performance for the eavesdropper. 
The open-loop prediction estimate and error covariance matrix are given by:
\begin{equation}\label{EQN_Open_Loop_Estimation}
x^{op}_k=\expe{ x_{k}}=0,\quad P^{op}_k=\var{x_k},
\end{equation}
with the covariance obeying the Lyapunov recursion: 
\begin{equation}\label{EQN_Open_loop_covariance}
P^{op}_k=AP^{op}_{k-1}A'+Q,\,P^{op}_{0}=\Sigma_{0}
\end{equation}
For the unstable states, where the open-loop prediction mmse explodes to infinity, we require the eavesdropper's mmse to explode as well.
For the stable states, we require the eavesdropper's mmse to converge to the open-loop one. 
\begin{definition}[Perfect Secrecy]\label{DEF_perfect_secrecy}
	Given system~\eqref{EQN_system} and channel model~\eqref{EQN_channel_model}, a coding scheme achieves perfect secrecy if and only if all of the following hold:
	\renewcommand{\labelenumi}{(\roman{enumi})}
	\begin{enumerate}
		\item the user's performance is optimal: \begin{equation}\label{EQN_optimal_estimation_user}
		\hat{x}_{u,k}=x_k,\,\text{when }\gamma_{u,k}=1.
		\end{equation}
		\item the eavesdropper's mmse for the unstable states grows unbounded with probability one:
		\begin{equation}\label{EQN_Definition_Perfect_Secrecy_1}
		[P_{k}]_{ii}\stackrel{a.s.}{\rightarrow}\infty,\text{ for }i=1,\dots,n_u
		\end{equation}
		\item the eavesdropper's mmse for the stable states converges to the open-loop prediction one with probability one:	\begin{equation}\label{EQN_Definition_Perfect_Secrecy_2}
		[P_{k}-P^{op}_k]_{ii}\stackrel{a.s.}{\rightarrow}0,\text{ for }i=n_u+1,\dots,n
		\end{equation}
		where  $\stackrel{a.s.}{\rightarrow}$ denotes almost sure convergence   as $k\rightarrow \infty$.
	\end{enumerate}
\end{definition}

The confidentiality requirements against the eavesdropper are only with respect to the current state and we do not consider guarantees for the batch state estimation error.
Contrary to our previous works~\cite{tsiamis2017codes,tsiamis2017codes_stable}, we have a unified problem formulation for general linear systems. This unification was not possible before; in~\cite{tsiamis2017codes}, we could only protect the unstable part of the state, while in~\cite{tsiamis2017codes_stable} the analysis was limited to stable systems. 
In the following section, we present a new code construction, based on a new analysis from the information matrix viewpoint, which is more general and solves the previous limitations.

\section{Coding scheme}\label{Section_Solution}
In this section, we first present State-Secrecy Codes for general linear systems. Then, we prove that they lead to perfect secrecy.
The sensor encodes and transmits the current state $x_k$ as a weighted state difference of the form $x_k-L^{k-t_k}x_{t_k}$, where $L$ is a carefully designed matrix.
State $x_{t_k}$, also called the \emph{reference state} of the encoded message, is the most recent state received at the user's end. The sensor and the user can agree on it via acknowledgment signals. Hence, the user can recover the current state by adding $L^{k-t_k}x_{t_k}$.  

However, if the eavesdropper fails to intercept $x_{t_{k}}$, then she cannot exactly decode neither $x_k$ nor the future packets. Any uncertainty about the reference state, $x_{t_k}$, gets amplified by $L^{k-t_k}$ when the eavesdropper tries to decode the current packet $x_k-L^{k-t_k}x_{t_k}$ to obtain $x_k$. This also obstructs the eavesdropper from decoding future packets, since the next reference state will depend on the current reference state $x_{t_{k}}$ and so on.
Missing just one $x_{t_k}$ triggers a chain reaction effect where the eavesdropper's mmse starts behaving as the open-loop prediction one and perfect secrecy is achieved. For this reason, we call this event \emph{critical}.
\begin{definition}[Critical event]\label{DEF_critical_event}
	A critical event occurs at time $k$ if the user receives the packet, while the eavesdropper fails to intercept it: $\gamma_{u,k}=1,\,\gamma_{k}=0$
	\hfill $\diamond$
\end{definition}

Let us now formally present the coding scheme.  
We define the \emph{reference time}  $t_{k}$ to be the time of the most recent successful reception at the user before $k$:
\begin{equation}\label{EQN_reference_time}
t_{k}=\max\set{t:\:0\le t<k,\,\gamma_{u,t}=1}.
\end{equation}
When the set $\set{t:\:0\le t< k,\, \gamma_{u,t}=1}$ is empty (before the first successful transmission), we use $t_{k}=-1$, $x_{-1}=0$.

	The code construction is based on the open-loop \emph{information matrix}, defined as the respective covariance inverse:
	\begin{equation}\label{EQN_information_matrix_open_loop}
	Y^{op}_{k}=(P_k^{op})^{-1}.
	\end{equation}
	It is well defined and positive definite since $Q,\Sigma_{0}\succ 0$ in~\eqref{EQN_Open_loop_covariance}.
	Since $A$ is unstable in general, $P^{op}_k$ will not converge. However, the information matrix $Y^{op}_k$ converges to a steady state matrix $Y_{\infty}$ as the following proposition shows.	
	\begin{proposition}\label{Prop_open_loop_converges}
Consider system~\eqref{EQN_system}. The open-loop prediction information matrix $Y^{op}_k$, defined in~\eqref{EQN_information_matrix_open_loop}, converges:
		\begin{equation}\label{EQN_Information_Steady}
		Y^{op}_k\rightarrow Y_{\infty}=\matr{{cc}0&0\\0&P^{-1}_{s,\infty}}.
		\end{equation}
		Matrix $P_{s,\infty}$ is the unique positive definite solution of
		\begin{equation}\label{EQN_Lyapunov_Stable}
		P_{s,\infty}=A_sP_{s,\infty}A_s'+Q_s,
		\end{equation}
		and is the limit of $P^{op}_{s,k}\in\REAL^{n_s\times n_s}$, the part of $P_{k}^{op}$ corresponding to the stable states.\hfill $\diamond$
	\end{proposition}
We can now introduce the proposed coding scheme.
\begin{definition}[State-Secrecy Codes]\label{DEF_name_codes}
	Given system \eqref{EQN_system}, a State-Secrecy Code applies the following linear operation: \begin{equation} \label{EQN_Coding_Scheme}
	\y{k}=x_k-L^{k-t_k}x_{t_k},\text{ with } L=A+Q(A')^{-1}Y_{\infty},
	\end{equation}
	where $t_k$ is the reference time as defined in \eqref{EQN_reference_time} and $Y_{\infty}$ is the steady-state information matrix~\eqref{EQN_Information_Steady}. \hfill $\diamond$
\end{definition}

The implementation of the scheme is described in Algorithm~\ref{ALG_coding}.
Notice that the code uses information about the model of the dynamical system. The intuition behind the form of $L$ is explained in Remarks~\ref{REM_artificial_dynamics},~\ref{REM_Weighting_Matrix_Form}. 
The next theorem proves that State-Secrecy codes achieve perfect secrecy if the critical event occurs at least once.

\begin{theorem}[Perfect secrecy]\label{THM_perfect_secrecy} 
	Consider system \eqref{EQN_system}, with channel model \eqref{EQN_channel_model} and coding scheme \eqref{EQN_Coding_Scheme}. 
	If the critical event occurs at least once:
	\begin{equation}\label{EQN_theorem_condition}
	\mathbb{P}( \gamma_{u,k}=1,\,\gamma_{k}=0,\text{ for some }k\ge 0 )=1, 
	\end{equation}
	then perfect secrecy is achieved; the user's estimation is optimal and satisfies~\eqref{EQN_optimal_estimation_user}, the eavesdropper's mmse behaves asymptotically as the open-loop one satisfying~\eqref{EQN_Definition_Perfect_Secrecy_1}, \eqref{EQN_Definition_Perfect_Secrecy_2}.
	\hfill $\diamond$
\end{theorem}  
\begin{algorithm}[!t]
	\caption{State-Secrecy Code}
	\label{ALG_coding}
	\begin{algorithmic}[1] {}
		\Require $A$, $Q$, $x_k$ for all $k\ge 0$.
		\Ensure Encoded signals $\y{k}$, for all $k\ge 0$.
		\Statex Let $t$ represent the time of user's most recent message.
		\State Compute $Y_{\infty}$ as in \eqref{EQN_Information_Steady} \State Set $L=A+Q(A')^{-1}Y_{\infty}$
		\State Initialize $t=-1$, $x_{-1}=0$
		\For{$k=0,1,\dots$ } 
		\State{Transmit $\y{k}=x_{k}-L^{k-t}x_{t}$}
		\If{Acknowledgment received} $t=k$ \EndIf
		\EndFor
	\end{algorithmic}
\end{algorithm}

Condition~\eqref{EQN_theorem_condition} for perfect secrecy is minimal since it only requires the critical event to occur once. For most channels of practical interest the critical event occurs not only once but infinitely often, e.g. when the outcomes are i.i.d. (see--Remark~1 in~\cite{tsiamis2017codes}). 
Even if the critical event never occurs naturally, we can force it by employing additional codes, i.e.,  encryption~\cite{Cryptography}, only at $k=0$. Then letting a simple State-Secrecy Code take over for $k>0$ leads to perfect secrecy. 
\begin{remark}[Comparison with previous codes]
From \eqref{EQN_Information_Steady}, \eqref{EQN_Coding_Scheme}, the weighting matrix $L$ can be rewritten as:
\begin{equation}\label{EQN_Coding_Scheme_Alternative}
L=\matr{{cc}A_{u}&Q_{12}(A'_s)^{-1}P^{-1}_{s,\infty}\\0&A_s+Q_{s}(A'_s)^{-1}P^{-1}_{s,\infty}}
\end{equation}
If the system has only unstable modes, then we recover the coding scheme in~\cite{tsiamis2017codes} with $L=A_u$. If the system is stable then we recover the coding scheme in~\cite{tsiamis2017codes_stable}, since from~\eqref{EQN_Lyapunov_Stable}:
\begin{equation}\label{EQN_Coding_Scheme_Alternative_Stable}
A_s+Q_{s}(A'_s)^{-1}P^{-1}_{s,\infty}=P_{s,\infty}(A'_s)^{-1}P^{-1}_{s,\infty}
\end{equation}
The current scheme is not just a diagonal combination of the codes in \cite{tsiamis2017codes,tsiamis2017codes_stable}. We  have the additional cross term $Q_{12}(A'_s)^{-1}P^{-1}_{s,\infty}$, which is necessary to achieve perfect secrecy, as illustrated in Section~\ref{Section_Simulations}.\hfill $\diamond$
\end{remark}

In the remainder, we study the eavesdropper's estimation performance via bounds on its mmse covariance and the respective information matrix. 
The former satisfies a nonlinear Riccati recursion (see Lemma~\ref{LEM_covariance_lower_bound}) while the latter follows a stable Lyapunov recursion, which is easier to analyze (see Lemma~\ref{LEM_covariance_convergence}). Hence, we can show that the information matrix bound converges to the open-loop prediction one. Then, Theorem~\ref{THM_perfect_secrecy} follows as a consequence.

Suppose that the critical event occurs at some time $k_0$. Then, we can establish the following lower bound $\bar{P}_{k}$ on the eavesdropper's mmse covariance.
\begin{lemma}[MMSE Bound]\label{LEM_covariance_lower_bound}
	Consider system~\eqref{EQN_system} channel model~\eqref{EQN_channel_model} and coding scheme~\eqref{EQN_Coding_Scheme}. If the critical event:
	\begin{equation}
	\event=\set{\gamma_{u,k_0}=1,\,\gamma_{k_0}=0}\label{EQN_event_B}
	\end{equation}
	occurs at some time $k_0\ge 0$, then with probability one:
	\begin{equation}\label{EQN_covariance_lower_bound}
	P_k\succeq \bar{P}_k,\text{ for }k\ge k_0,\text{ in }\event
	\end{equation}
	where $\bar{P}_{k}$ satisfies the Riccati equation for $k\ge k_0$:
	\begin{equation}\label{EQN_covariance_Riccati}
	\bar{P}_{k+1}=L\bar{P}_{k}L'-L\bar{P}_{k}H'\paren{H\bar{P}_{k}H'+Q}^{-1}H\bar{P}_{k}L',
	\end{equation}
	with $H=A-L$ and $\bar{P}_{k_0}=P_{k_0}\succ 0$. \hfill $\diamond$
\end{lemma}
The lower bound $\bar{P}_k$
is equal to the true covariance $P_k$ when the eavesdropper intercepts all packets after $k_0$:\[P_k=\bar{P}_k\text{, conditioned on } \event\cap\set{\gamma_{k}=1\text{ for all }k>k_0}.\] 
 Since $(L,H)$ is not detectable we can not use the classical tools of Kalman filter to study~\eqref{EQN_covariance_Riccati}. Nonetheless, it is easier to work with the information matrix version of $\bar{P}_k$, which satisfies a Lyapunov recursion with stable dynamics.

\begin{lemma}[Convergence properties]\label{LEM_covariance_convergence}
Consider the Riccati recursion~\eqref{EQN_covariance_Riccati}, for $k\ge k_0$, and some $k_0\ge 0$, with $H=A-L$ and $L$ as in~\eqref{EQN_Coding_Scheme}. If $\bar{P}_{k_0}\succ 0$ then:
\begin{enumerate}[a)]
	\item The information matrix bound:
\begin{equation}\label{EQN_information_matrix_lower_bound}
\bar{Y}_k=\paren{\bar{P}_k}^{-1}, \text{ for }k\ge k_0.
\end{equation} is well defined and satisfies the Lyapunov recursion:
\begin{equation}\label{EQN_Lyapunov_lower_bound}
\bar{Y}_{k+1}=(L')^{-1}\bar{Y}_k L^{-1}+(L')^{-1}H'Q^{-1} H L^{-1}
\end{equation}
\item Matrix $\bar{Y}_k$  converges to the open-loop one:
\begin{equation}\label{EQN_Information_Matrix_Convergence}
\bar{Y}_{k}\rightarrow Y_{\infty},
\end{equation}
where $Y_{\infty}$ is defined in~\eqref{EQN_Information_Steady}.
	\item For the unstable states the lower-bound mmse diverges:
\begin{equation}\label{EQN_Unstable_Divergence}
[\bar{P}_{k}]_{ii}\ge c_i\abs{\lambda_{i}}^{2(k-k_0)},\text{ for }i=1,\dots,n_u,
\end{equation}
where $c_i>0$ are positive constants.
	\item For the stable states, the lower-bound mmse is at least equal to the open-loop one asymptotically:
	\begin{equation}\label{EQN_Stable_Convergence}
\liminf_{k\rightarrow \infty}[\bar{P}_{k}-P^{op}_k]_{ii}\ge  0,\text{ for }i=n_u+1,\dots, n
	\end{equation}
	where $P_{s,\infty}$ is defined in~\eqref{EQN_Lyapunov_Stable}.
	 \hfill $\diamond$
\end{enumerate}
\end{lemma}
Equation~\eqref{EQN_Lyapunov_lower_bound} holds for arbitrary $L$ and is central to the analysis of state-secrecy codes as it captures all of their convergence properties. 
It is also linear in $\bar{Y}_k$ and easier to analyze than~\eqref{EQN_covariance_Riccati}.
The results c), d) of the above lemma essentially prove perfect secrecy for the  covariance $\bar{P}_k$. Then, Theorem~\ref{THM_perfect_secrecy} follows since $P_k$ will eventually be lower bounded by $\bar{P}_k$ for $k\ge k_0$ and some $k_0$. The following remarks provide  intuition about the codes.
\begin{remark}[Artificial Dynamics]\label{REM_artificial_dynamics}
From~\eqref{EQN_Coding_Scheme_Alternative}, and~\eqref{EQN_Coding_Scheme_Alternative_Stable} the eigenvalues of $L$ are:\begin{align}
\mathrm{eig}\paren{L}
&=\mathrm{eig}\paren{A_u}\cup \mathrm{eig}\paren{A_s^{-1}},\label{EQN_Unstable_Eigenvalues}
\end{align}
i.e., all the eigenvalues are outside the unit circle. Hence, in~\eqref{EQN_covariance_Riccati} matrix $L$ imposes explosive unstable artificial dynamics to the eavesdropper's estimation scheme; any prior uncertainty about $x_{k}$ is amplified by $L$, when the eavesdropper attempts to decode $z_{k+1}$.
By selecting a purely unstable matrix $L$ we force the eavesdropper's information to be upper bounded over time by $\bar{Y}_k$ in~\eqref{EQN_Lyapunov_lower_bound} since the Lyapunov recursion will converge; if $L$ had stable eigenvalues, the eavesdropper's information for some states would grow unbounded.
\end{remark}

\begin{remark}[Rate optimality]\label{REM_Weighting_Matrix_Form}
	Matrix $L$ shows up in the open-loop information matrix recursion. 
	If we write the open-loop covariance recursion~\eqref{EQN_Open_loop_covariance} in terms of the information matrix $Y^{op}_k$ (by applying the inversion Lemma~\ref{LEM_inversion_lemma} (see Appendix)), we obtain the Riccati recursion:
	\begin{equation*}
	Y^{op}_{k+1}=FY^{op}_{k}F'-FY^{op}_{k}F'\paren{FY^{op}_{k}F'+W}^{-1}FY^{op}_{k}F',
	\end{equation*}
	where $F=(A')^{-1}$ and $W=Q^{-1}$.
	This is equivalent to:
	\begin{equation}\label{EQN_Open_Loop_Riccati}
	Y^{op}_{k+1}=\paren{F-K_kF}Y^{op}_{k}\paren{F-K_kF}'+K_kWK_k',
	\end{equation}
	with $K_k=FY^{op}_{k}F'\paren{FY^{op}_{k}F'+W}^{-1}$. 
	From the proof of Proposition~\ref{Prop_open_loop_converges} and~\eqref{EQN_Inverse_of_L} in the Appendix it follows that
	\[F-K_kF\rightarrow (L')^{-1}, \]
	 while 
	 \[K_k\rightarrow -(L')^{-1}H'.\]
	  Hence, asymptotically the open-loop information matrix recursion~\eqref{EQN_Open_Loop_Riccati} matches the eavesdropper's information matrix recursion~\eqref{EQN_Lyapunov_lower_bound}.
	In this respect, the convergence rate in~\eqref{EQN_Lyapunov_lower_bound} is asymptotically optimal.   \hfill $\diamond$
\end{remark}

\section{Extensions and Discussion}\label{Section_Discussion}

\noindent\textbf{General matrix $\boldsymbol{A}$. } 
We can deal with singular $A$ by slightly perturbing the zero eigenvalues. For example, if $\lambda_i=0$ is a simple eigenvalue corresponding to state $x_i$, we can define $\bar{A}=A+\delta e_{i}e_i'$, where $e_i\in\REAL^{n}$ is the $i$-th canonical vector (all elements $0$ and the $i$-th element is $1$) and $\delta>0$ is a small constant to be designed. We then replace $A$ with $\bar{A}$ in the code construction. If $\delta$ is small enough, then the eavesdropper's $x_i$-mmse error will not converge exactly to the open-loop one, but it will remain close to it.
If matrix $A$ has eigenvalues on the unit circle, we can treat them as being in the unstable part in~\eqref{EQN_Coding_Scheme_Alternative}. A result similar to~\eqref{EQN_Unstable_Divergence} holds. Suppose that $\lambda_{j}=1$. Then, one occurrence of the critical event will only guarantee that the mmse for the marginally stable state is lower bounded, i.e. $[\bar{P}_{k}]_{jj}\ge c_i$. Instead of once, we need the critical event to occur infinitely often in order to achieve unbounded mmse for state $x_j$. The formal analysis of both cases is left for future work.

\noindent\textbf{Unreliable acknowledgments. } 
Suppose that the reverse channel is also a packet dropping one. 
Let $\gamma_{a,k}\in\set{0,1}$ denote the reverse channel outcome at time $k$. If $\gamma_{a,k}=1$, then the sensor successfully receives the respective acknowledgment, otherwise it does not. 
Then, we can redefine the reference time to be:
\[
\bar{t}_{k}=\max\set{t:\:0\le t<k,\,\gamma_{u,t}\gamma_{a,t}=1},
\]
where we require both the packet and the acknowledgment to be successfully transmitted to update the reference time. To make sure that the user knows $\bar{t}_k$, the sensor should also transmit $\bar{t}_k$ at every time step. If we define $\bar{\gamma}_{u,k}=\gamma_{u,k}\gamma_{a,k}$ the results of this paper still hold if we replace $\gamma_{u,k}$ with $\bar{\gamma}_{u,k}$.

\section{Simulations}\label{Section_Simulations}
\begin{figure}[t] \centering{
		\includegraphics[scale=0.48]{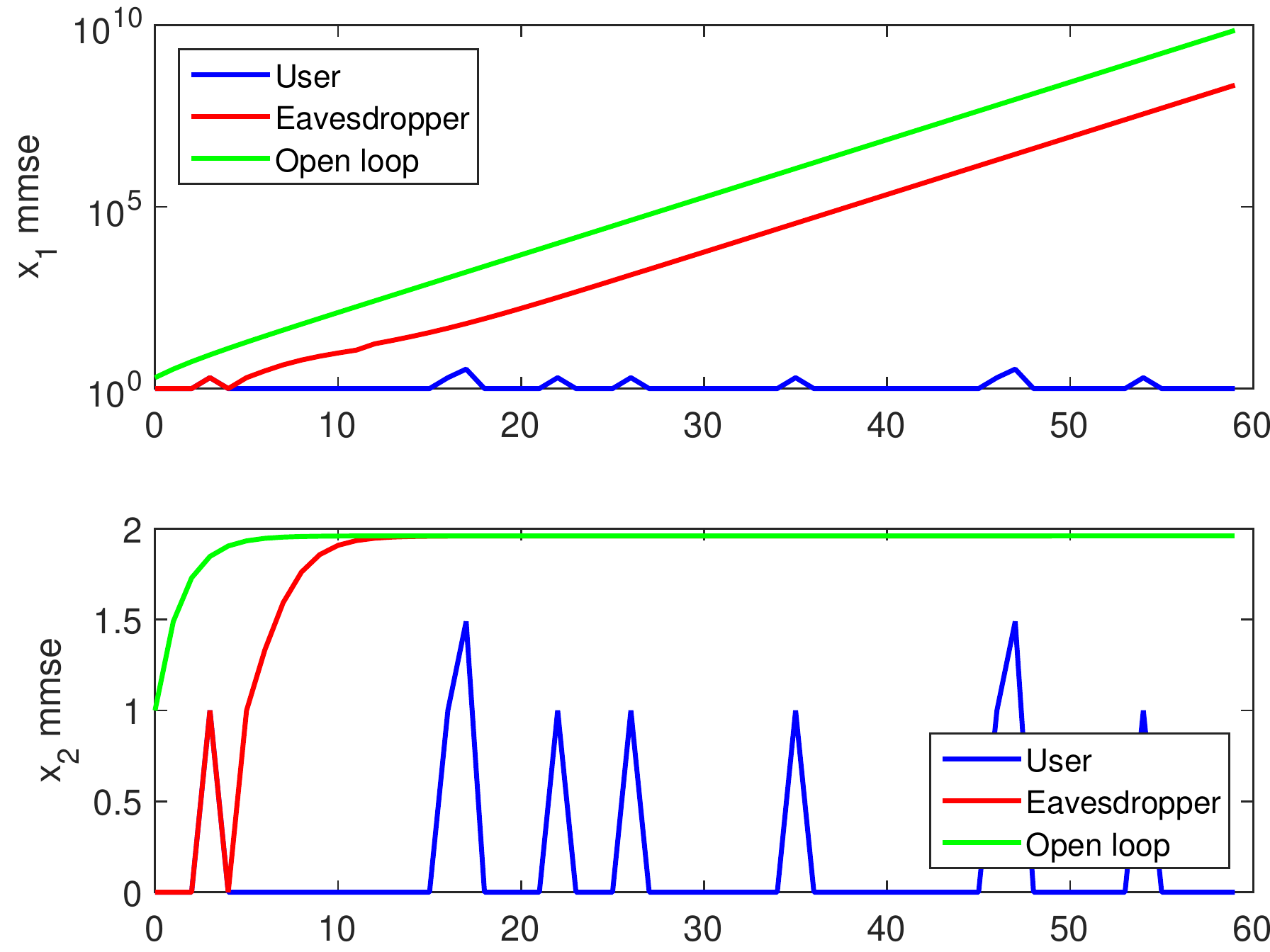}}   
	\caption{ We compare the eavesdropper's, user's and open-loop mmse for the states $x_1$ and $x_2$. For the log-plot we used function $\log\paren{x+1}$ instead of $\log x$. For this random sample of channel outcomes, the critical event occurs at time $k_0=5$. Then, the eavesdropper's mmse error for the unstable state $x_1$ starts diverging, while the mmse error for the stable state $x_2$ starts converging to the open-loop prediction one. The user has zero error at the successful reception times.}
	\label{Figure_performance}
\end{figure}
We illustrate the performance of State-Secrecy Codes via numerical simulations. 
The system under consideration has state matrix $A=\matr{{cc}1.2& 0\\0& 0.7}$ and noise covariance matrices $\Sigma_0=Q=\matr{{cc}1& 0.8\\0.8& 1}$. For the channel model, we assume that the channel outcomes are independent across time and stationary with probabilities $P(\gamma_{u,k}=i,\gamma_{k}=j)=p_{ij}$, for $i,\,j\in\set{0,1}$. 
The assumed values are $p_{11}=0.7$, $p_{01}=p_{10}=p_{00}=0.1$.
 Since the user can decode all signals, we used the formula:
\begin{equation*}
P_{u,k}=
\left\{\begin{aligned}
&0&&\text{ if }\gamma_{u,k}=1\\
&AP_{u,k-1}A'+Q&&\text{ if }\gamma_{u,k}=0
\end{aligned}\right.
\end{equation*}
For the estimation scheme of the eavesdropper see \cite{tsiamis2017codes_stable} or \cite{tsiamisJournal} for more details.

In Figure~\ref{Figure_performance}, we plot the user's and eavesdropper's mmse  for the states $x_1$, $x_2$, i.e. the diagonal elements of the matrices $P_{u,k}$, $P_{k}$. We compare them to the open-loop prediction error $P^{op}_k$ defined in~\eqref{EQN_Open_Loop_Estimation}. The eavesdropper's mmse error for the unstable state $x_1$ starts diverging after the first critical event occurs at time $k_0=5$. Meanwhile, the mmse error for the stable state $x_2$ starts converging to the open-loop prediction one. 
The user can decode all received messages and has zero error at the times of successful reception.

\begin{figure}[t] \centering{
		\includegraphics[scale=0.48]{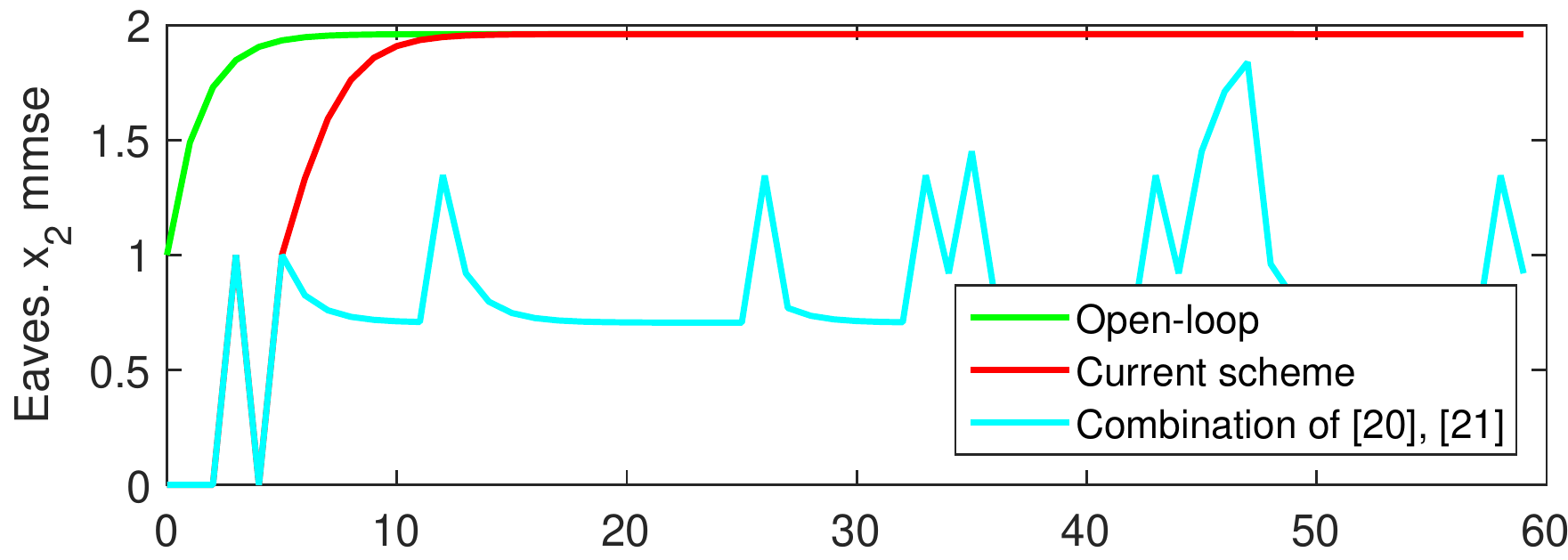}}   
	\caption{ We compare our coding scheme with the diagonal combination of those in~\cite{tsiamis2017codes,tsiamis2017codes_stable}. The comparison is made for the eavesdropper's $x_2-$mmse error. The diagonal combination fails to achieve perfect secrecy.}
	\label{Figure_comparison}
\end{figure}

In Figure~\ref{Figure_comparison} we compare scheme~\eqref{EQN_Coding_Scheme}, with the diagonal combination of the schemes in~\cite{tsiamis2017codes,tsiamis2017codes_stable}, i.e. if we use
\[\bar{L}=\matr{{cc}A_u&0\\0&P_{s,\infty}(A_s')^{-1}P^{-1}_{s,\infty}}.\] The comparison is made for the stable state $x_2$. The eavesdropper's mmse error under $\bar{L}$ fails to converge to the open-loop one. This shows that the cross-term $Q_{12}(A_s')^{-1}(P_{s,\infty})^{-1}$ in~\eqref{EQN_Coding_Scheme_Alternative} is necessary for perfect secrecy.
\section{Conclusion}\label{Section_Conclusion}
By exploiting the  model of the dynamical system, the channel randomness, and the artificial dynamics, State-Secrecy Codes offer strong confidentiality guarantees for the current state with minimal computational cost and no communication overhead. With just a single occurrence of the critical event, the eavesdropper's information starts decreasing with asymptotically the same rate as the open-loop information. 
In future work, the codes should be adapted to the case of output measurements and closed loop systems, i.e., when the user is a controller. Further studies should be made about the information leakage regarding the past states.  Making our scheme more robust against active eavesdroppers is also another future direction.

\appendix 
\subsection*{Inversion lemmas}
The following two lemmas are from chapter~0 in~\cite{horn2012matrix}.
\begin{lemma}[Inversion Lemma~\cite{horn2012matrix}]\label{LEM_inversion_lemma}
	Let $B,C,U,V$ be matrices of conformable sizes with $B,C$ invertible:
	\begin{equation*}
(B+UCV)^{-1}=B^{-1}-B^{-1}U(C^{-1}+VB^{-1}U)^{-1}VB^{-1}.
	\end{equation*}
\end{lemma}

\begin{lemma}[Block Inversion Lemma~\cite{horn2012matrix}]\label{LEM_block_inversion_lemma}
	Let $B=\matr{{cc}B_{11}&B_{12}\\B_{21}&B_{22}}$ be an invertible matrix. Let $C=B^{-1}$ with the same partition. Assuming the involved inverses exist:
	\begin{align*}
	C_{22}=(B_{22}-B_{21}B_{11}^{-1}B_{12})^{-1},\,
	C_{21}=-C_{22}B_{21}B_{11}^{-1}.
	\end{align*}
\end{lemma}
The second identity follows from $CB=I.$
\subsection*{Proof of Proposition~\ref{Prop_open_loop_converges}}
The open-loop information matrix satisfies the Riccati equation~\eqref{EQN_Open_Loop_Riccati}, which does not meet the stabilizability condition. Thus, we leverage the results of~\cite{chan1984convergence} for non-stabilizable systems. The proof proceeds in two steps. First, we show that $Y_{\infty}=\matr{{cc}0&0\\0&P_{s,\infty}^{-1}}$ is a stabilizing solution (see below) to the algebraic version of~\eqref{EQN_Open_Loop_Riccati}. Then, we prove convergence.

\textit{Part A: stabilizing solution. }There are two conditions for $Y_{\infty}$ to be a stabilizing solution~\cite{chan1984convergence}:  
\begin{enumerate}[i)]
	\item  Matrix $Y_{\infty}$ is a fixed-point of~\eqref{EQN_Open_Loop_Riccati}
	\item All eigenvalues of 
	$F-K_{\infty}F$ are inside the unit circle, where $
	K_{\infty}=FY_{\infty}F'\paren{FY_{\infty}F'+W}^{-1}.
	$
\end{enumerate}
Lengthy algebra gives:
\[
K_{\infty}=\matr{{cc}0&0\\ P^{-1}_{s,\infty}Q_{12}'&P^{-1}_{s,\infty}Q_s},
\]
where we used~\eqref{EQN_Lyapunov_Stable}, Lemma~\ref{LEM_block_inversion_lemma} two times (one to express $Q_{s}^{-1}$ and one for $\paren{FY_{\infty}F'+W}^{-1}$), and $Q^{-1}_sQ_{12}'=-W_{21}W_u^{-1}$ (follows from $QQ^{-1}=I$). This also implies:
\[
F-K_{\infty}F=\matr{{cc}(A_u')^{-1}&0\\-P_{s,\infty}^{-1}Q_{12}'(A_u')^{-1}&P_{s,\infty}^{-1}A_s P_{s,\infty}}.
\]
To verify i), we compute:
\begin{align*}
FY_{\infty}F'-K_{\infty}FY_{\infty}F'=\matr{{cc}0&0\\0&D},
\end{align*}
with
$
D=(A')^{-1}P_{s,\infty}^{-1}A^{-1}-P^{-1}_{s,\infty}Q_s(A')^{-1}P_{s,\infty}^{-1}A_s^{-1}.
$
If we replace $Q_s$ with $P_{s,\infty}-A_sP_{s,\infty}A_s'$, we verify $D=P_{s,\infty}^{-1}$. This shows that $Y_{\infty}$ is a fixed-point of~\eqref{EQN_Open_Loop_Riccati}.
To verify ii), notice that \[\mathrm{eig}(F-K_{\infty}F)=\mathrm{eig}(A_u^{-1})\cup \mathrm{eig}(A_s).\] Thus, from Assumption~\ref{ASSUM_system} the eigenvalues of $F-K_{\infty}F$ lie inside the unit circle. Thus, $Y_{\infty}$ is a stabilizing solution.

\textit{Part B: convergence}
 We use Theorem~4.2 of~\cite{chan1984convergence}. Since the pair $(F,F)$ is observable ($F$ is invertible), $F$ has no poles on the unit circle and $Y^{op}_{0}=\Sigma_0^{-1}\succ 0$, matrix $Y^{op}_k$ converges exponentially fast to the unique stabilizing fixed-point of~\eqref{EQN_Open_Loop_Riccati}, i.e., $Y_{\infty}$ from part A.
 Finally, $P_{s,k}^{op}$ satisfies the Lyapunov recursion:
 \[
 P_{s,k}^{op}=A_s P_{s,k-1}^{op}A_s'+Q_s.
 \]
 Since $A_s$ is stable, $P_{s,k}^{op}$ converges to $P_{s,\infty}$. \hfill $\QED$
\subsection*{Proof of Theorem~\ref{THM_perfect_secrecy}}
The user can always decode the messages, thus,  condition~\eqref{EQN_optimal_estimation_user} of perfect secrecy is true.
To prove the remaining conditions assume that the critical event $\event=\set{\gamma_{u,k_0}=1,\gamma_{k_0}=0}$ occurs for some $k_0$. 
From Lemmas~\ref{LEM_covariance_lower_bound}, \ref{LEM_covariance_convergence}, 
 conditioned on $\event$:
 \begin{align*}
[P_{k}]_{ii}&\ge c_i \abs{\lambda_i}^{2\paren{k-k_0}}\rightarrow \infty,\text{ for }i=1,\dots,n_u\\
\liminf_{k\rightarrow \infty}&[P_k-P_k^{op}]_{ii}\ge 0,\text{ for }i=n_u+1,\dots,n
 \end{align*}
From Lemma~3 in~\cite{tsiamis2017codes_stable}, we also have $P_k\preceq P_k^{op}$. Thus:
\[
\lim_{k\rightarrow \infty}[P_k-P_k^{op}]_{ii}= 0,\text{ in }\event\text{ for }i=n_u+1,\dots,n.
\]
Since conditions~\eqref{EQN_Definition_Perfect_Secrecy_1}, \eqref{EQN_Definition_Perfect_Secrecy_2} hold in $\event$ for every $k_0\ge 0$, from hypothesis~\eqref{EQN_theorem_condition} they hold with probability one.\hfill $\QED$
\subsection*{Proof of Lemma~\ref{LEM_covariance_lower_bound}}
The proof of $P_k\succeq \bar{P}_k$, $P_{k_0}=\bar{P}_{k_0}$ can be found in the proof of Theorem~1 in ~\cite{tsiamis2017codes_stable}. The proof of~\eqref{EQN_covariance_Riccati} is the same as the proof of Lemma~1 in~\cite{tsiamis2017codes_stable}. From the same proof we can also deduce that $\bar{P}_{k_0}=\Sigma_0\succ 0$ or $\bar{P}_{k_0}\succeq Q\succ 0$. \hfill $\QED$

\subsection*{Proof of Lemma~\ref{LEM_covariance_convergence}}
 \textit{Proof of a).} Equation~\eqref{EQN_Lyapunov_lower_bound} follows from a direct application of the inversion Lemma~\ref{LEM_inversion_lemma} to~\eqref{EQN_covariance_Riccati}. By Lemma~\ref{LEM_covariance_lower_bound}, we have $\bar{P}_{k_0}\succ 0$, thus, $\bar{Y}_{k_0}\succ 0$ and is well defined.
By induction, from~\eqref{EQN_Lyapunov_lower_bound}, matrix $\bar{Y}_k\succ 0$ is well defined for all $k\ge k_0$.

 \textit{Proof of b).} By~\eqref{EQN_Unstable_Eigenvalues} matrix $L$ is purely unstable, which implies $L^{-1}$ has all eigenvalues inside the unit circle. Hence, matrix $\bar{Y}_k$ converges to the unique positive semi-definite fixed point of the Lyapunov recursion~\eqref{EQN_Lyapunov_lower_bound}. Thus, it is sufficient to show that $Y_{\infty}$ satisfies the
 Lyapunov equation:
  \begin{equation}\label{EQN_Proof_Lyapunov_target}
Y_{\infty}=(L')^{-1}Y_{\infty} L^{-1}+(L')^{-1}H'Q^{-1} H L^{-1}.
 \end{equation}
First, from~\eqref{EQN_Coding_Scheme_Alternative} and~\eqref{EQN_Coding_Scheme_Alternative_Stable} the inverse of the block lower triangular matrix $L'$ is:
\begin{equation}\label{EQN_Inverse_of_L}
(L')^{-1}=\matr{{cc}(A_u')^{-1}&0\\-P_{s,\infty}^{-1}Q_{12}'(A_u')^{-1}&P^{-1}_{s,\infty}A_sP_{s,\infty}}.
\end{equation}
The first term in the right-hand side of~\eqref{EQN_Proof_Lyapunov_target} is:
\begin{equation}\label{EQN_Proof_Lyapunov_first_term}
(L')^{-1}Y_{\infty} L^{-1}=\matr{{cc}0&0\\0&P^{-1}_{s,\infty}A_sP_{s,\infty}A_s'P^{-1}_{s,\infty}}
 \end{equation}
Next, we compute the second term of~\eqref{EQN_Proof_Lyapunov_target}. We have:
\[
(L')^{-1}H'=\matr{{cc}0&0\\-P^{-1}_{s,\infty}Q_{12}&-P^{-1}_{s,\infty}Q_{s}}.
\]
After some algebra and using the definition of the inverse $QQ^{-1}=I$, we obtain:
\begin{equation}\label{EQN_Proof_Lyapunov_second_term}
(L')^{-1}H'Q^{-1}HL^{-1}=\matr{{cc}0&0\\0&P^{-1}_{s,\infty}Q_sP^{-1}_{s,\infty}}.
\end{equation}
From~\eqref{EQN_Information_Steady}, \eqref{EQN_Proof_Lyapunov_target}, \eqref{EQN_Proof_Lyapunov_first_term}, \eqref{EQN_Proof_Lyapunov_second_term}, we only need to check if the nonzero elements are equal or:
\[
P_{s,\infty}^{-1}=P^{-1}_{s,\infty}A_sP_{s,\infty}A_s'P^{-1}_{s,\infty}+P^{-1}_{s,\infty}Q_sP^{-1}_{s,\infty}.
\]
But this follows from~\eqref{EQN_Lyapunov_Stable} if we multiply with $P_{s,\infty}$ from both sides. This completes the proof of b).

\textit{Proof of c).} Define the operator:
\begin{equation}\label{EQN_Operator}
g(X)=LXL'-LXH'(HXH'+Q)^{-1}HXL',
\end{equation}
which is increasing with respect to positive semidefinite comparison~(Lemma~1c in~\cite{Sinopoli2004Kalman}).

Assume that state $x_i$ corresponds to a simple real unstable eigenvalue $\lambda_i$ for some $i\in\set{1,\dots,n_u}$. Define $e_i$ to be the $i$-th canonical vector ($i$-th element is $1$ and the remaining are $0$). Recall that $A$ is in real Jordan form, thus $A_{ii}=\lambda_i$. Since $\bar{P}_{k_0}\succ 0$, we have $\bar{P}_{k_0}\succeq c_i e_ie_i'$, where $c_i=\lambda_{\min}(\bar{P}_{k_0})>0$ is the minimum eigenvalue of $\bar{P}_{k_0}$. From~\eqref{EQN_Coding_Scheme_Alternative} it follows that: 
\begin{equation}\label{EQN_output_matrix}
H=A-L=\matr{{cc}0 &-Q_{12}(A')^{-1}P^{-1}_{s,\infty}\\0&-Q_{s}(A')^{-1}P^{-1}_{s,\infty}}.
\end{equation}
This implies $He_i=0$, for $i=1,\dots,n_u$.
 or 
\[g(c_i e_ie_i')=c_i Le_ie_i'L'=c_i \lambda_i^{2} e_ie_i',\]
By monotonicity of $g$, we have:
\[
\bar{P}_{k_0+1}=g( \bar{P}_{k_0})\succeq g(c_i e_ie_i')=c_i\lambda_i^{2} e_ie_i'.
\]
Repeating, it follows by induction that:
\[
\bar{P}_{k}\succeq c_i \lambda_i^{2\paren{k-k_0}} e_ie_i',\,k\ge k_0.
\]
Thus, $[\bar{P}_{k}]_{ii}=e_i'\bar{P}_{k} e_i\ge c_i \lambda_i^{2\paren{k-k_0}} $. The case of complex eigenvalues or multiple eigenvalues is similar and, thus, omitted. One can start from $\bar{P}_{k_0}\succeq \lambda_{\min}(\bar{P}_{k_0}) \paren{e_ie_i'+\dots+e_{i+p}e_{i+p}'}$, where $p$ is the dimension of the corresponding real Jordan block.

 \textit{Proof of d).}
Denote the block partition of matrix $\bar{P}_{k}$ as:
\[
\bar{P}_{k}=\matr{{cc}\bar{P}_{u,k}&\bar{P}_{12,k}\\\bar{P}_{12,k}'&\bar{P}_{s,k}}.
\]
The partitions of matrix $\bar{Y}_k$, $P^{op}_k$ are defined similarly.
From the block inversion Lemma~\ref{LEM_block_inversion_lemma}, we have
 \[
\bar{Y}^{-1}_{s,k}=\bar{P}_{s,k}-\bar{P}_{12,k}'\bar{P}_{u,k}^{-1}\bar{P}_{12,k}\preceq \bar{P}_{s,k}.
\]
For the diagonal elements, the above inequality implies:
$
[\bar{Y}^{-1}_{s,k}]_{jj}\le [\bar{P}_{s,k}]_{jj},\text{ for }j=1,\dots,n_s.
$
or
\[
\liminf_{k\rightarrow \infty} [\bar{P}_{s,k}-P^{op}_{s,k} ]_{jj}\ge \lim_{k\rightarrow\infty}[\bar{Y}^{-1}_{s,k}-P^{op}_{s,k} ]_{jj}=0,
\]
where the last equality follows from Proposition~\ref{Prop_open_loop_converges} and b). This completes the proof of d) since by definition of $\bar{P}_{s,k}$, $P^{op}_{s,k}$ the above inequality is the same as~\eqref{EQN_Stable_Convergence}.

\bibliographystyle{IEEEtran}
\bibliography{IEEEabrv,Paper_wireless_secrecy}
\end{document}